\documentclass[12pt]{article}

\usepackage{graphicx}
\usepackage{hyperref}

\usepackage{amsmath, amsthm, amssymb}

\begin{document}
\title{Introducing the Computable Universe\thanks{Based in the introduction to \emph{A Computable Universe}, published by World Scientific Publishing Company and the Imperial College Press, 2012. See \url{http://www.mathrix.org/experimentalAIT/ComputationNature.htm}}}
\author{Hector Zenil\\Department of Computer Science, The University of Sheffield, UK;\\ IHPST (Paris 1 Panth\'eon-Sorbonne/ENS Ulm/CNRS), France;\\ \& Wolfram Science Group, Wolfram Research, USA.}

\date{}

\maketitle

\begin{abstract}
Some contemporary views of the universe assume information and computation to be key in understanding and explaining the basic structure underpinning physical reality. We introduce the Computable Universe exploring some of the basic arguments giving foundation to these visions. We will focus on the algorithmic and quantum aspects, and how these may fit and support the computable universe hypothesis.
\end{abstract}

\section{Understanding Computation \& Exploring Nature As Computation}

Since the days of Newton \index{Newton, I.} and Leibniz\index{Leibniz, G.}, scientists have been constructing elaborate views of the world. In the past century, quantum mechanics\index{quantum mechanics} revolutionised our understanding of physical reality at atomic scales, while general relativity\index{general relativity} did the same for our understanding of reality at large scales. 

Some contemporary world views approach objects and physical laws in terms of information and computation \cite{zenilinfo}, to which they assign ultimate responsibility for the complexity in our world, including responsibility for complex mechanisms and phenomena such as life. In this view, the universe and the things in it are seen as computing themselves. Our computers do no more than re-program a part of the universe to make it compute what we want it to compute. 

Some authors have extended the definition of computation to physical objects and physical processes at different levels of physical reality, ranging from the digital to the quantum. Most of the leading thinkers involved in this effort are contributors to \emph{A Computable Universe} \cite{zenillibro}, including some who oppose the (digital) approach, preferring to advance their own. 

The computational/informational view (sometimes identified as \textit{computationalism}\index{computationalism}) is rooted in pioneering thinking by authors such as John A. Wheeler\index{Wheeler, J.A.} and Richard Feynman. A quotation from Feynman's\index{Feynman, R.} Messenger Lectures, delivered at Cornell University in 1964, distills his sense that nature most likely operates at a very simple level, despite seeming complex to us, and marks a shift to understanding physics in terms of digital information.

\begin{quotation}
It always bothers me that, according to the laws as we understand them today, it takes a computing machine an infinite number of logical operations to figure out what goes on in no matter how tiny a region of space, and no matter how tiny a region of time ... So I have often made the hypothesis that ultimately physics will not require a mathematical statement, that in the end the machinery will be revealed, and the laws will turn out to be simple, like the chequerboard with all its apparent complexities.
\end{quotation}

His view was probably influenced by his thesis advisor John A. Wheeler, the coiner of the phrase ``it from bit'', suggesting that information constitutes the most fundamental level of physical reality  \cite{wheeler}. Another pioneer was Konrad Zuse, a new edition of whose Calculating Space\index{Calculating Space} (\textit{Rechnender Raum})\index{Rechnender Raum} we are pleased to be able to publish in \emph{A Computable Universe} \cite{zenillibro}. We've put great effort into translating into modern \LaTeX \text{ } the scanned version of the original translation commissioned by Ed Fredkin and published by MIT. Fredkin is himself another pioneer, having founded the field that is today known as \textit{digital physics}, and is a contributor to \emph{A Computable Universe} \cite{zenillibro}. Stephen Wolfram, a student of Feynman, has also been building upon this view, spearheading a paradigm shift facilitated by today's availability of increasingly greater and cheaper computational resources. Even more recently, some modern theories of physics under development have been attempting to unify quantum mechanics and general relativity on the basis of information ('t Hooft, Susskind, Smolin), a development represented in the work of some contributors to \emph{A Computable Universe} \cite{zenillibro}. 

These contemporary views of a computational universe are also deeply related, via the concept of information, to a contemporary field of mathematical research called algorithmic information theory (AIT). AIT researchers think that the true nature of nature can only be unveiled by studying the notion of randomness (Greg Chaitin\index{Chaitin, G.}, Leonid Levin\index{Levin, L.}, Cris Calude\index{Calude, C.}). A few papers are devoted to this topic, including two from Chaitin and Calude \footnote{Where author names are provided without a reference, the authors in question are contributors to \emph{A Computable Universe} \cite{zenillibro}.}.

\subsection{What is computation? How does nature compute?}

Zuse suggested early on that the world was possibly the result of deterministic digital computation, in particular a cellular automaton. Ed Fredkin\index{Fredkin, Ed} would later develop the idea further (see his contribution to \emph{A Computable Universe} \cite{zenillibro}). In his minimalistic approach, Wolfram argues that the world may turn out to be the result of very simple rules, perhaps even a single one, from which the apparent complexity we see around us emerges. If the universe is a computable one, we could just run a universal Turing computer\index{universal Turing computer} on every possible program to generate not only our own universe but every possible one, as Wolfram\index{Wolfram, S.} and Schmidhuber\index{Schmidhuber, J.} have suggested (both are contributors to \emph{A Computable Universe} \cite{zenillibro}). The question would then be how to distinguish our universe from any other. Wolfram has pointed out that there will be some universes that are obviously different from ours, and many others that may look very similar, in which case we can ask whether ours will turn out to be special or uncommon in any sense, whether, for example, it would rank among the first in terms of description size\index{description size}, i.e. be among those having the shortest description. If it is simple enough, AIT would then suggest that it is also frequent, the result of many programs generating the same universe. This is a conclusion based on algorithmic probability, which describes the distribution of patterns, and establishes a strong connection to algorithmic complexity.

According to Schmidhuber's approach, it would seem that a computer generating every possible universe would necessarily have to be several times larger than the universe itself. But if the programs are short enough, instead of running every program a step at a time, beginning with the smallest and proceeding in increasing order of size,  it may be possible to start with some plausible universes and run them for longer times, as Wolfram has suggested, checking each program and allowing those that are apparently complex to unfold (and eventually leading to the physical properties of our universe).

There are also those who believe that nature not only performs digital computation (as is the view of Zuse, Fredkin or Wolfram, all contributors to \emph{A Computable Universe} \cite{zenillibro}), but is itself the result of quantum computation (Lloyd\index{Lloyd, S.}, Deutsch\index{Deutsch, D.} and Cabello\index{Cabello, A.}, also authors of chapters in \emph{A Computable Universe} \cite{zenillibro}). According to them, the world would in the last instance be rooted in physics, particularly quantum mechanics, and would reflect the properties of elementary particles and fundamental forces. Lloyd, for example, asks how many bits there are in the universe, offering an interesting calculation according to which, given the properties of quantum particles, the universe cannot be rendered in a description shorter than itself--simply because every elementary particle would need to be simulated by another elementary particle. A computer to simulate the universe would therefore need to be the size of the universe, and would require the energy of the actual universe, hence making it undistinguishable from a (quantum) computer,\index{quantum computer} the computer and the computed being in perfect correspondence. 

If the goal is not to describe with the greatest accuracy our universe in the state it is in, one may ask whether a universe of similar complexity would require a very complicated description. The discussion seems therefore to revolve around the possible description of the universe, whether it can be written in bits or in qubits, whether it can be shorter than the universe or not. If digital information\index{digital information} underlies the quantum, however, it may turn out that the shortest description of the universe would be much shorter than the universe itself, contrary to the views of, for example, Lloyd, Deutsch, Cabello or Calude\index{Calude, C.} et al., and could then be compressed into a simple short computer program, as Zuse, Wolfram, Fredkin, Bolognesi\index{Bolognesi, T.} and I myself believe. 

In fact, one can think of the goal of digital physics as a minimal model describing the universe, equivalent to the goal of physics in its quest for a unified theory, a descriptive formula governing all forces and particles yet to be discovered, if any such remain. Whether such a program exists is an open question, just as it is an open question whether there is a theory of everything (ToE)\index{theory of everything}. But it is no longer an open question whether such unification can be achieved for very large portions of physics. It is a fait accompli in fields such as gravitation and movement, electricity and magnetism, electromagnetism and most electronuclear forces\index{electronuclear forces}, areas of physics that today model with frightening accuracy large portions of nature using simple laws than can be programmed in a computer and can in  principle (and often de facto) provide perfect prescriptions and predictions about the world. In fact we have physical laws and computer programs for pretty much everything; what we lack is a single theory that encompasses all other theories.

\section{The algorithmic approach}

If the world is in fact not a digital computer, it could nevertheless behave like one. Thus whether or not it is a digital computer, one could test whether the output of processes in the world resembles the output that one would expect from running a program. There are many laws that computers may follow. Or if you prefer, they follow a specific subset of physical laws related to information processing, notably the distribution of patterns described by algorithmic probability. 

From my point of view, information can only exist in our world if it is carried by a process; every bit has to have a corresponding physical carrier. Even though this carrier is not matter, it takes the form of an interaction between components of matter--an atom interacting with another atom, or a particle interacting with another particle. At the lowest level, however, the most elementary particles, just like single bits, carry no information (the Shannon entropy of a single bit is 0 because one cannot implement a communication channel of 1 bit only, 1 and 0 having the same possible information content if taken in isolation). Isolated particles may have no causal history, being memoryless when isolated from external interaction. When particles\index{elementary particles} interact with other particles they appear to be linking themselves to a causal network and seem to be forced to define a value as a result of this interaction (e.g. a measurement\index{measurement problem}). What surprises us about the quantum world is precisely its lack of apparent causality, which we see everywhere else and are so used to. But it is the interaction and its causal history that carries all the memory of the system, with the new bit appearing as if it had been defined at random because our theories of quantum mechanics only provide probability amplitudes. Linking a bit to the causal network may seem tantamount to producing a correlation of measurements between seemingly disconnected parts of space, while in fact they may have always already been connected, if the world were taken as deterministic (a view implicit in so-called \textit{hidden variables}\index{hidden variables} models). 

Levin's universal distribution\index{Levin's universal distribution}  \cite{kirchherr} based on algorithmic probability describes expected output frequencies in relation to their complexity. A process that produces a string $s$ with a program $p$ when executed on a universal Turing machine $T$ has probability $Pr_T(s)=2^{-|p|}$ where $|p|$ is the length of the program $p$. 

The coding theorem\index{coding theorem}  \cite{cover,calude} connects the frequency $Pr(s)$ with which a string $s$ is produced to its algorithmic complexity $C(s)$. The so-called \textit{semi-measure} $m$ has also the remarkable property of dominating  $Pr_T$ for any universal Turing machine\index{Turing machine} $T$. Roughly speaking, $m(s)$ establishes that if there are many long descriptions of a certain string, then there is also a short description with low algorithmic complexity $C(s)$, that is $m(s) \approx 2^{-C(s)}$. As neither $C(s)$ nor $m(s)$ is computable, no program can exist which takes a string $s$ as input and produces $m(s)$ as output. However, we have proven  \cite{delahayezenil} that numerical approximations are possible and that reasonable numerical evaluations produce reasonable results.
\index{Kolmogorov complexity}\index{Kolmogorov-Chaitin complexity}\index{algorithmic complexity}\index{Algorithmic Information Theory}
 Just as strings can be produced by programs, we may ask after the probability of a certain outcome from a certain natural phenomenon, if the phenomenon, just like a computing machine, is a process rather than a random event. If no other information about the phenomenon is assumed, we can see whether $m(s)$ says anything about a distribution of possible outcomes in the real world  \cite{zenilalgo}. In a world of computable processes, $m(s)$ would indicate the probability that a natural phenomenon produces a particular outcome and tell us how often a certain pattern would occur.
\index{$\pi$}

Consider an unknown operation generating a binary string of length $k$ bits. If the method is uniformly random, the probability of finding a particular string $s$ is exactly $2^{-k}$, the same as for any other string of length $k$, which is equivalent to the chances of picking the digits of $\pi$ at random. However, data (just like $\pi$--largely present in nature, for example, in the form of common processes relating to curves) are usually produced not at random but by a specific process (in the case of $\pi$ there are many ways to produce it; there are even physical phenomena that lead to it). This is where $m(s)$ may be relevant to calculating the probability of physical processes.

\subsection{Information and structure in living organisms}

Biology has witnessed a transformation during the last century, beginning with Mendel's\index{Mendel, G.} discoveries regarding the transfer of certain traits in pea plants, a phenomenon amounting to an information transfer between generations. Later rediscovered, his laws would lay the foundation for what is today the modern science of genetics, which has established that living organisms store and bequeath information comprising instructions for their full development encoded--as Watson\index{Watson, J.} and Crick\index{Crick, F.} discovered--in ribonucleic and nucleic acids. The code of life is digital; two bits per base pair are needed to encode the DNA. 

Rules determining the way DNA replicates\index{DNA replication} may be algorithmic in nature, like those governing other types of physical phenomena, leading us to sometimes discover strong similarities in their pattern distribution. Processes of DNA are relatively simple. A subset of purely digital operations can match their operations with computational ones, operations such as joining, copying, partitioning, complementation, trimming, or replacing. Which implies that layer upon layer of the code of life has been built up over billions of years in a deep algorithmic process with its own characteristic rules, making processes like protein folding\index{protein folding} appear highly complex to us. If we ignore the algorithm of protein folding, there is no reason to think that protein folding cannot be carried out by a (deterministic) machine (whether in polynomial time or not). 

In the case of structures whose final state is certain (whether a folded protein or the division of a cell), the solution is either the result of an algorithm or of a random process. It is unlikely that protein folding is random because an incorrectly folded protein would obviously cause disease. One of the most important properties of life is robustness, which we think may be explained by algorithmic probability  \cite{zenilmarshall}.

 If computation is the driving force producing structure in the world, one can use it as a basis for manipulating the direction of complexity. Think of the problem of putting together the right chemical elements for life. If we expect life to emerge out of chemicals, expecting them to produce, say, the digits of the mathematical constant $\pi$, the chances of this happening are ridiculously low. How has nature produced organisms of such complexity? Wolfram, for example, thinks that nature mines what he calls the computational universe of possible programs. The concept of Darwinian\index{Darwin, C.}\index{Darwinian evolution} evolution may lead us to assume that whatever the processes are that give rise to the forms we see in biological systems, they must have been fully shaped by natural selection\index{natural selection}. But there is strong evidence (e.g.~Ref.~\cite{wolfram}) that nature samples programs, that in some way nature may be disposing what computation proposes. If the same question about pattern production is now asked in the context of computation, the probability of chemicals self-assembling\index{self-assembling} to produce the digits of $\pi$ by chance is substantially larger, because one needs to find a program producing $\pi$ rather than the digits themselves. Programs producing $\pi$ are infinitely shorter than the number of digits in $\pi$. 

Patterns in nature could also be the result of a similar way of shortcutting pattern formation, and the cause of the structure we see in nature. Patterns would not repeat by mimicking themselves; they would recur because they are the result of a simple program producing the same pattern over and over, with the frequency dictated by algorithmic probability. Should the pattern change, it would mean that the rule has changed too, but as algorithmic probability predicts, if the program remains short, the chances of its  producing a pattern of low algorithmic complexity, perhaps even the same one that it was producing before a mutation, is exponentially large, as compared to the chances of its producing random-looking patterns. It would make better sense then to think of natural selection as taking place at the level of programs, in order to preserve the fundamental property of resilience in living structures, i.e., robustness. Patterns are robust then because they are the result of a short, concise rule.

 On the other hand, if one were free to select random patterns rather than random programs, the result would be rather random-looking. Yet this seems not to be what happens, which is why we don't see sudden changes in patterns in nature, even if we do experience some apparent randomness. A legitimate question to ask is the role of this apparent randomness in biology, whether it is what drives biological speciation based on mutation. Is this randomness\index{apparent randomness} only apparent? Ultimately, is the world more random than structured (see, e.g. Calude. Another proponent of this view is McAllister\index{McAllister, J.}  \cite{mcallister}). 

How could the world be a simple computer program? If every physical process is itself computable, Turing showed that there is a single machine capable of running all of them. In other words, one can build a single program out of many. Of course, claiming that processes or that the world itself may be a computer program doesn't mean that they actually are, or that they behave like Turing machines. We should not let ourselves be fooled into thinking that the proposition that nature is computable means that the universe is a Turing machine. Obviously there are natural processes that are definitely not like Turing machines. A Turing machine is an oversimplification of the concept of (digital) computation. The human brain, for example, is a very different object to a Turing machine, despite the parallels between them. The question of whether (all) physical processes in the human brain can be carried out by a Turing machine shouldn't be taken as suggesting that the brain is a Turing machine. In \emph{A Computable Universe} \cite{zenillibro}, Szudzik\index{Szudzik, M.} and Hutter\index{Hutter, M.} provide precise mathematical formalisms for the computable universe hypothesis\index{computable universe hypothesis}. 

But if the world is dominated by computable processes\index{computable processes}, of the kind that can be carried out by digital computation, then much of the structure of the world may be credited to computation alone, because computation would follow the universal distribution of the frequency of patterns that algorithmic probability describes. The fact that, their limitations notwithstanding, scientific models can describe much of the world, and that such models are computable, strongly suggests that though nature may do more than compute, to a large extent its activity does amount to (Turing) computation. In the course of history we have managed to make nature do what we wanted it to do. The process began with simple tools, ultimately leading to computers that can be programmed to perform all kinds of calculations for us, and used to make other devices do all kinds of work, serving as tools. If nature does more than just compute, we know that it nevertheless can compute, and that it does so very well. One argument that may be advanced against this view is the inability of scientific models to predict with arbitrary precision. But their non-linearity by no means implies indeterminism or uncomputability. As Wolfram has shown, even the simplest computer programs (such as the elementary cellular automaton (ECA) rule 30\index{Wolfram rule 30}) are very difficult, if not impossible to understand well enough to be able to predict the manner of their unfolding--despite its overwhelming simplicity.

We don't have many reasons, perhaps none at all, to believe that there is anything inherent in nature that is not Turing computable, even if unpredictable. The fact that we are always able to replace theories with more encompassing and more accurate ones can be taken as a computable process in itself, in the sense of Turing's very definition of computability. Using $\pi$ again, $\pi$ is a computable number because a Turing machine can compute any arbitrary number of the digital expansion of $\pi$. So even if we are not able to exhibit and calculate the full expansion of $\pi$ in any base, the fact that it is computable means that we can always provide a better approximation of it. For any digit with index $i$ in the expansion of $\pi$ we can always calculate $i+1$ digits of $\pi$. $\pi$ is also random-looking and in many respects unpredictable, as we cannot tell, without making a calculation, what an arbitrary digit will be (not even what the chances are, as it is believed to be normal, every digit has equal probability). Even if there are formulae (BBP  \cite{bbp}) to calculate arbitrary digits of $\pi$ in bases $2^n$ that do not require the calculation of the previous digits, the result of the calculation of a digit of $\pi$ doesn't make it any easier to recognise. The digits of $\pi$ in all the bases in which it has been studied looks random by all current statistical tests, and although there is no proof of its normality, it is believed to be so, meaning that it has a property of a true random number, viz. that it contains every possible pattern, though it is not itself a pattern, being random. 

As for unknown formulae for $\pi$, for ECA rule 30, one cannot rule out the possibility of formulae of the same or completely different type that nonetheless permit rapid computation of individual digits of the evolution of these apparent random systems.

\section{Determinism from Quantum Mechanics?}

Unlike the apparent randomness from previous section from $\pi$ or ECA rule 30, in quantum mechanics events seem to happen for no apparent reason---for example, the time at which a particle decays or the position of an electron collapsing from entanglement. Surprisingly enough, the most serious idea that the universe is digital comes from quantum mechanics (with its roots in a long tradition beginning with the ancient Greek concept of atomism). It was Max Plank, who had been trying to understand the emission of radiation from heated objects, who discovered the quantisation phenomenon of radiation. Quantum mechanics is in effect thought of as being digital, but it is at the same time analogue. An electron can be in many places simultaneously, covering an extended region of space. So quantum mechanics hasn't settled the issue once and for all, because particles themselves behave as waves, thus preserving a duality between digital and analogue descriptions of the most fundamental building blocks of the universe (for a discussion of this topic see Ref.~\cite{zenilfqxi}). This is not a question of philosophy, nor even of mathematics; it is a question of physics, with a definite answer. If we zoomed in far enough into reality, and if the world is digital, we should be able to see pixels and bits. However, if the world is not digital, we would always be able to zoom in without being able to prove or disprove any possibility. A third answer is that it is neither one nor the other, implying that it is actually both, just as the standard interpretation of quantum mechanics would have it.

 There is a tendency, however, that has favoured a discretisation of models (for some discussion of this matter see Ref.~\cite{zenilinfo}. Electricity was first thought to be continuous, but then the electron was discovered. Likewise  with the photon, and gravitation is today approached in the same way. It would seem that we keep going from a continuous theory to an atomic one, ending up with an unsatisfactory answer (particle duality leading to an explanation of electricity in terms of electrons, which themselves behave both as particles and waves). In this process, we have come to realise that information plays an important role, as what many of these theoretical mergers actually do to objects or phenomena is to divest them of certain characteristics once believed to be exclusively theirs and attribute them to the composite object or phenomenon of which they are now thought to form a part.

 The development of quantum mechanics early in the last century prompted physicists to radically rethink the concepts they used to describe the world. No current account of what information may be can be considered complete that does not take into account the interpretations of quantum mechanics. Classical systems comply with the criteria of what may be identified as local-realism, that is, that the results of measurements of a system localised in space-time are fully determined by properties inherent in that system (its physical reality\index{physical reality}), and cannot be instantaneously influenced by a distant event (locality). In other words, locality prohibits any influences between events in space-like separated regions, while realism claims that all measurement outcomes depend on pre-existing properties of objects that are independent of measurement. Quantum phenomena, however, seem fundamentally non-local in the relativistic sense, as one is forced to invoke a particular frame of reference to give a sense to the statement that measurement of a particle happens first, and that its result immediately affects the state of a second particle, even if placed beyond the limit allowed for information exchange between the two by the theory of relativity. 

As discovered by Albert Einstein\index{Einstein, A.}, Boris Podolsky\index{Podolsky, B.} and Nathan Rosen\index{Rosen, N.}, \index{EPR experiment}quantum mechanics predicts strong correlations between measurements carried out on two particles in an entangled state. It is tempting to interpret these correlations as the result of shared properties determined at the time of their initial interaction and then assimilated into each particle. Bell's formulation of an inequality  \cite{bell} made it possible to settle the debate by performing an experiment to test the inequalities by showing that hidden-variables theories based on the joint assumption of locality and realism are at variance with the statistical predictions of quantum mechanics. 

Ever since, physicists have undertaken experiments to test quantum reality. The first of these was John Bell\index{Bell, J.} himself, who showed that if one assumed that particles were correlated, in the sense that measuring the properties of one tells you the properties of the other, then there was an inequality which described the maximum possible correlation in a classical world. Later refinements of Bell's inequality tests have continuously converged, closing both the locality loophole on the one hand, and the detection loophole on the other. Therefore quantum foundation scientists think that it is reasonable to consider the violation of local realism a well established fact. 

For the Bell experiment to validate quantum reality\index{quantum reality}, one would need to design it to cover all possible open ends or loopholes. Could it be that there were some biases induced by the Bell experiments\index{Bell experiments} that resulted in particles that were more correlated having a greater chance of being the ones measured? That would explain the violation. Alain Aspect\index{Aspect, A.}  \cite{aspect} designed an experiment to test and rule out this possibility. It involved the use of several detectors measuring a large number of photon pairs in order to obtain statistically significant results. After several experiments, the community reached a point where it was convinced that Aspect's experiments\index{Aspect's experiments} ruled out the possibility that Bell's experiment was statistically biased, measuring only correlated particles. Could it be that this and other experiments are still fooling us, making us believe that quantum reality behaves in a certain way when it actually does not? 

Aspect helped to rule out the problem of dealing with local bias in the measurement of correlated particles, but what if something were influencing the experiment by communicating properties between particles? One way to rule out this possibility is by ensuring a distance between the correlated particles, a distance sufficient to guarantee, by the speed of light, that if the two measurements at the two ends were fast enough, nothing could travel to communicate anything about one particle to the other. Remember that the assumption here is that even when the two particles are correlated because they come from a single source, they could not be correlated beyond what the Bell inequality establishes, and so if the correlation is greater than that, it means that there is an eerie entanglement between quantum particles that leads one of them to know or change its properties once the other has been measured.

 If two of these particles are close enough to communicate with each other, the violation of Bell's inequality is easily explained. One should guarantee that information cannot therefore be disseminated without a physical carrier, a carrier of matter which cannot travel faster than the speed of light. And so by placing two particles far enough apart, one can try to rule out this possibility (again if it is not the case that there is an underlying layer of information, a hidden-variables reality storing or communicating the measured values). The availability of highly efficient sources of entangled particles (with the emergence of laser technology) has allowed quantum scientists to perform \textit{Aspect experiments} with a distance large enough to apparently close this loophole, preventing particles from communicating with each other without having to assume that they have done so faster than the speed of light, which general relativity renders out of the question. 

This can only work, however, if the settings of the detectors are changed every time, given that previous experiments may have influenced them or that one detector may have figured out the settings of the other. For in order to guarantee the validity of the resulting measurement of correlation, there must not be any correlated measurements at the outset. Weihs et al. \index{Weihs, H.} \index{Zeilinger, A.} \cite{zeilinger} managed to separate two particles by 400 meters, giving them 1.3 microseconds to switch the detectors' settings randomly, and the maximum 5 nanoseconds between the measurements guaranteed that, at the speed of light, no information could possibly be transmitted between the two ends. Even as some loopholes in tests of quantum mechanics are closed, others may open up (e.g. \textit{collapse locality}) and no test has yet encompassed all loopholes at once. The local-realistic hidden variables theory preferred by Einstein\index{Einstein, A.} may not be a viable description of the world, but there are other loopholes to close, particularly one that is assumed and that ultimately leads to a circular argument.

It turns out that the aforementioned attempt at closing a loophole assumes the ability to randomly change the settings of the detectors on each side, by keeping their random generators far enough apart to ensure that the choice of one random generator does not influence the other. This is a loophole that no experimental physicist thinks is worth trying to close, because there is little, if anything, to test. It turns out that all tests of Bell's inequality\index{Bell's inequality} assume that one has the freedom to choose the detectors' settings and that such freedom has to be indeterministic for the interpretation of the results of the experiment to work. But in practice one is obliged to choose between two options: either to use a pseudo-random generator, which means that the source of randomness is actually not random but deterministic and therefore potentially correlated to the other detectors' settings, or that one is free to choose from a set of non-commuting operators\index{non-commuting operators}, i.e. free will\index{free will} and true (indeterministic) randomness\index{true randomness} to begin with when preparing the experiment. 

If the universe runs deterministically, however, there is nothing to explain and all quantum strangeness makes immediately sense. This is in agreement with what Bell himself acknowledged as a possibility: if quantum mechanics is completely deterministic, then even the experimenter's decision to measure certain components of the spins is entirely pre-determined, so that the observer could not possibly have decided to measure anything other than what he does in fact measure. In other words, the correlations would have a trivial explanation, because it is trivial to think of correlated states at all levels, given that everything in the universe would indeed be the effect of a previous state and would therefore simply share a common origin. By analogy, think of similar sets of chromosomes between siblings. Sharing genes explains correlations in their eye colour or other features, without having to introduce other convoluted explanations. That everything in the universe is determined since the very beginning is of course a strong commitment to make in order to make sense of quantum experiments, but it is no doubt it is much simpler by most, if not all means, than  current interpretations of quantum mechanics. 

 There must be a reason to try to force an interpretation of quantum mechanics where the only possible outcome is indeterministic randomness, even if it means assuming none other than indeterministic randomness. When the theory of quantum mechanics was launched in the works of Plank\index{Plank, M.}, Einstein and Bohr\index{Bohr, N.}, it was clear that classical physics could not explain small-scale phenomena such as the interaction between particles. The theory of quantum mechanics could, but at great cost, since one had to be ready to make all kinds of concessions to certain phenomena incompatible with classical physics. For Bohr, these concessions constituted reality at the level of the quantum world, while Einstein thought the weirdness of quantum mechanics meant that the theory was incomplete, that in the end it couldn't possibly be as incompatible as it was with the familiar reality we inhabit. He was convinced that an yet undiscovered theory of hidden variables would be capable of explaining some, if not all, of these strange phenomena. 

\begin{quote}
The logical conclusion one can draw from the violation of local realism is that at least one of its assumptions fails. But one could consider the breakdown of other assumptions that are implicit in our reasoning leading to the inequality. These may include a violation of the limit imposed by relativity theory or a world that is completely deterministic  \cite{bell}.
\end{quote}

This loophole is often said to be avoided by quantum physicists concerned with quantum reality, who assume that no test of complete determinism can be performed. Nevertheless, pseudo-randomness has the obvious particularity that when its generating process is repeated, the result is exactly the same. If this is the case, then this very same phenomenon would also occur in quantum mechanics. The difficulty is perhaps that the formalism of quantum mechanics only gives us probability amplitudes, but what we may be witnessing in quantum mechanics in the form of unexpected correlations is simply actually an indication of determinism that we have tried too hard to explain otherwise.

For if quantum indeterministic randomness is possible, it still remains to be explained how that randomness may have any impact on classical mechanics (one of, if not the most important discrepancy with relativity). For example, at the quantum scale, what governs particles is time-reversible, and all physical laws preserve such time-reversal at the quantum level. Yet this time-reversal has apparently no impact on our macroscopic reality, suggesting that there may be a causal disconnect here. This may also be the case with quantum randomness, if it exists. It may be that quantum randomness does not have any impact beyond the quantum scale. Think of entanglement. Thus far it has been quite difficult to take advantage of quantum superposition in the classical world, and this may turn out to be an unsurpassable barrier. For when an entangled particle interacts with the macroscopic world, its state collapses to a single determined one and the quantum state is lost, which would appear to unavoidably set classical mechanics apart from quantum mechanics.


Producing random bits in a discrete universe, where all events are the cause and effect of other events, would actually be very expensive, if not produced for free by quantum mechanics. For one would need to devise a way to break the causal network--assuming that this were possible to begin with--to produce a random bit and yet leave the rest of the causal network untouched (otherwise we would see nothing but randomness, which is not the case).

\section{From String Theory to Bit String Theory}

On the other hand, information is taking on a fundamental character in modern physics, black holes, string theory, etc. The assumption that information is essential to explain our physical reality is not alien to modern physics. Many physicists have arrived at similar conclusions, giving the concept of symmetry (which may be seen as information or an abstract mathematical object) a foundational role, even in predicting the existence of new particles, for example. And so far this approach has been quite successful in many areas of physics. It is remarkable how a simple informational description of some fundamental properties of quantum mechanics fully describes quantum phenomena. Such descriptions specify how particles interact and how they are related to each other by symmetries.

 According to classical mechanics, randomness is apparent in the macroscopic world, but under the standard interpretation of quantum mechanics things are fundamentally different. The position that the history of the world is computationally reversible is compatible with the determinism imposed by classical mechanics. At the quantum scale, however, things seem otherwise. When one entangles a particle the particle seems to truly and irreversibly lose track of its previous state, and there is no way, even by reversing all operations, to recover it. This also happens with radioactive decay, where an atomic nucleus of an unstable atom loses energy by emitting ionising particles for which, according to quantum theory, it is impossible to predict the decay time, decay being equally likely to occur at any given time. 

Zeilinger claims that quantum randomness is intrinsically indeterministic, and that experiments violating Bell's inequality imply that some properties do not exist until measured  \cite{zeilinger2}. These claims are, however, based on a particular interpretation (if not speculation) of quantum mechanics, from which he leaps to conclusions by relying on various no-go or no-hidden-variables theorems proposed by people like von Neumann\index{von Neumann, J.}, Bell, Kochen and Specker\index{Kochen and Specker theorem}, which are supposed to show that quantum randomness is truly indeterministic. Zeilinger's position is a form of weak epistemological randomness, as he claims that one cannot observe less than one bit of information, and what is not known is then ``random".

My position resonates with the response to Zeilinger in  \cite{daumer}. But although I share with Daumer et al. the belief that Wheeler did not shed much light on the issue with his captivating but somehow rather obscure treatment of information as being related to, or as more fundamental that physics (his ``it from bit" dictum\index{it from bit}), I do not share their views as regards what's wrong with the informational content of quantum mechanics. As they point out, Wheeler's remarkable suggestion was that physics is only about information or that the physical world itself is information. 

In the words of J.A. Wheeler himself   \cite{wheeler}:

\begin{quote}
It is not unreasonable to imagine that information sits at the core of physics, just as it sits at the core of a computer. It from bit. Otherwise put, every \textit{it}---every particle, every field of force, even the space-time continuum itself---derives its function, its meaning, its very existence entirely---even if in some contexts indirectly---from the apparatus-elicited answers to yes-or-no questions, binary choices, bits.
\end{quote}

Wheeler, however, interjects a nuance related to meaning when introducing the observer (remember Wheeler is also credited with formulating the \textit{anthropic principle}). So in Wheeler's view, meaning is also subjective, which means that it is not a completely reductionist, trivial view of information content. 

\begin{quote}
`It from bit' symbolizes the idea that every item of the physical world has at bottom--a very deep bottom, in most instances--an immaterial source and explanation; that which we call reality arises in the last analysis from the posing of yes/no questions and the registering of equipment-evoked responses; in short, that all things physical are information-theoretic in origin and that this is a participatory universe.
\end{quote}

Wheeler's most pointed suggestion is that ``information" can't be defined in terms of ``matter" or ``energy" and that it may therefore be as or more fundamental than either ``matter" or ``energy", the most basic notions in physics. But a second reading also introduces the problem of information content, meaning, and the observer in his participatory universe.

\subsection{An algorithmic approach to the problem of fine tuning}

The anthropic principle can now have a plausible interpretation under this algorithmic approach, providing a (more reasonable) answer to the question of the apparent fine tuning of the universe, that is, the question of why the universe looks just right for accommodating everything in it. A single value changed in the equation would produce a universe where nothing would be possible, certainly not life. But if the universe is a computer program, parameters are not only coupled together but there may be many computer programs producing the same output, especially an output of low algorithmic complexity, that is, a structured universe rather than a random looking one--contra the position taken by Calude et al. in \emph{A Computable Universe} \cite{zenillibro}. The fact that parameters are coupled is also a reflection of a possible ToE, according to which everything comes into being as a result of a single physical law. Hence one may inquire into other possible ToEs that model well behaved universes where things seem just right, as they do in ours. In Wolfram's approach, for example, computer universes that are not trivial very quickly begin to look complicated enough for us to judge whether it obviously is or isn't our universe that's being modelled. This means that there is a threshold (captured in Wolfram's Principle of Computational Equivalence) where a universe may begin to look as if it had been fine tuned to accommodate all the structure it is capable of, without being at all special. On the contrary, as it is capable of structure it would have a low algorithmic complexity and therefore a high algorithmic probability. 

Following Wheeler, I think that the next level of unification (along the lines of the unification of other previously unrelated concepts in science, such as electricity and magnetism, light and electromagnetism, and energy and mass, to mention a few) will involve information and physics (and ultimately, as a consequence, computation and physics). 

Landauer wrote in 1996 (Quoted from Leff and Rex, pag. 335   \cite{demon}):

\begin{quote}
Information is not a disembodied abstract entity; it is always tied to a physical representation. It is represented by an engraving on a stone tablet, a spin, a charge, a hole in a punched card, a mark on paper, or some other equivalent. This links the handling of information to all the possibilities and restrictions of our real physical world, its laws of physics and its storehouse of available parts.
\end{quote}

Current attempts to unify modern physics, such as string theory and quantum gravity, depend on information encodings and on how much information is needed to describe something. In these models, it is information that imparts sense to the forces in the universe and to matter itself. In quantum gravity\index{quantum gravity} space and time are not fundamental; it is information that constitutes the most basic level of physical reality. That is, everything arises out of information. Relativity had already relegated space and time to this status, because there is nothing special about either dimension; it is just what happens in each that distinguishes space from time. Events taking place in them have only a subjective meaning, with information playing a fundamental role. In quantum gravity too, the exchange of information is fundamental. These theories are believed to introduce a compatibility between quantum mechanics and general relativity through the concept of information (in particular, \textit{maximum information}), connecting energy from quantum theory and energy from relativity theory, and finally bridging the two.

\subsection{Black holes as perfect data compressors}\index{Black holes}

Le\'o Szil\'ard \cite{szilard}\index{Szil\'ard, L.} and Rudolf Landauer \cite{landauer}\index{Landauer, R.} determined how a computer uses energy when processing information. It turns out that one can see computers as spending information only when erasing information. One can add or multiply or subtract bits without consuming or increasing the energy in the universe, as long as no information is \textit{erased} in the process. Landauer's principle\index{Landauer's principle} is derived from solid physical principles. Unfortunately, one has almost always to erase information, for almost everything. Erasing releases heat, which dissipates in the environment. The conclusion is that erasing is not a reversible operation. Erasing is not like subtracting or adding because subtraction and addition are reversible operations, while erasing entails a loss of local information, the heat released containing the lost information. The loss of information at the level of quantum mechanics is apparently of a different nature. 

It turns out that black holes play an important role in the informational view of our world given that the maximum information when bits are matched to photons is the maximal mass determined by the black hole limit. At first, black holes were thought to delete information from the universe; whatever fell into them would irreversibly be destroyed, despite the generalised assumption from thermodynamics that information in the universe is never destroyed (Bennett \cite{bennettenergy}). It was later acknowledged, however, that even black holes may conserve information in the form of a quantum phenomenon occurring at the event horizon by emitting small amounts of thermal radiation  \cite{hawking} from which the information in whatever fell into the black hole could be fully recovered, in principle. 

In fact, black hole formation can now be determined, in the context of what has been called the holographic principle\index{holographic principle} \cite{holo}, given by the number of bits of information equal to the surface of the black hole divided by four multiplied by the Planck area\index{Planck area} ($10^{-66} m^2$). Among several important consequences is that the maximum information pertaining to a black hole is finite, and that information pertaining to this physical object can be understood in terms of the information in two dimensions (its surface), for which reason the associated principle is called holographic---by analogy with the fact that the two-dimensional surface of an object contains information regarding its volume in three dimensions  \cite{bekenstein}\index{Bekenstein, J.}. This black-hole thermodynamics defines the maximum amount of information that can potentially be stored in a given finite region of space which has a finite amount of energy. The more information a black hole has, the larger its event horizon, and in no smaller place such information could be contained, as black holes as maximally larger when reaching the point of singularity. Black holes can be regarded thus as perfect data compressors.\footnote{An interesting question in connection to Kolmogorov complexity, given that Kolmogorov complexity is uncomputable, is how nature may have achieved a perfect data compressor, even though its contents and mechanisms may turn out to be inaccessible to us in practice.}


Quantum decoherence\index{quantum decoherence} means that the information an observer has about the probabilities of the different possible outcomes involves some specific physical degrees of freedom. But again, the measurement problem is just that the observer does not know the outcome of a quantum measurement, and as such quantum measurements appear not to be reversible, because entangling a particle again and performing the same experiment would lead to a different result, one that's discrepant from the previous measurement, not because it is random in the mathematical (algorithmic) sense, but in the same sense as we don't know, and we seem unable to predict the classical world, not because it is indeterministic, but because it is unpredictable. Even if there are subtle differences, it all gets down to a problem of epistemological nature, information at the end.
 
This goes to show how information may already explain some of the most fundamental processes in the universe at the edge of our understanding of the largest and the tiniest, and underscores the pertinence of asking seriously whether the universe can be fully explained in terms of information and computation, perhaps even suggesting ways to conciliate general relativity with quantum mechanics.



\end{document}